\documentclass[pra,amsmath,amssymb,preprint,byrevtex]{revtex4}

\usepackage{bm}
\usepackage{amssymb}
\usepackage{amsmath}
\usepackage{graphicx}

\newtheorem{lemma}{Lemma}

\newtheorem{theorem}{Theorem}

\newcommand{\tr}{\mathop{\mathrm{Tr}}\nolimits}

\begin{document}

%---------- FRONT MATTERS ---------------------------------------
%\preprint{}
\title{The Bloch Vector for $N$-Level Systems}
\author{Gen Kimura}
\email{gen@hep.phys.waseda.ac.jp}
\affiliation{Department of Physics, Waseda University, Tokyo 169--8555, Japan}
%\date[]{Apr 15, 2002}
\begin{abstract}
We determine the set of the Bloch vectors for $N$-level systems, generalizing the familiar Bloch ball in $2$-level systems. An origin of the structural difference from the Bloch ball in $2$-level systems is clarified.
\end{abstract}
%\pacs{03.65.-w, 03.65.bz}
\maketitle

%---------- INTRODUCTION ----------------------------------------
\section{Introduction}

The determination of a state on the basis of the actual measurements (experimental data) is important both for experimentalists and theoreticians. In classical physics, it is trivial because there is a one-to-one correspondence between the state and the actual measurement. On the other hand, in quantum mechanics, where a density matrix is used to describe the state, it is generally nontrivial to connect them \cite{ref:vonNeumann,ref:Pauli,ref:Des,ref:Weigert,ref:Peres1}.

%In $N$-level systems (especially in $2$-level systems), i

It is known that the Bloch vector (coherence vector) \cite{ref:Bloch,ref:HioeEberly,ref:Alicki,ref:Nielsen,ref:Mahler,ref:Jakobczyk,ref:Lendi,ref:Lendi2} gives one of the possible descriptions of $N$-level quantum state which meet the above requirement, because it is defined as a vector whose components are expectation values of some observables: For $N$-level systems, the number of observables that we need to identify the state is $N^2-1$\cite{ref:Weigert2}, since there are $N^2-1$ independent parameters of the density matrix $\hat{\rho}$ which is Hermitian and subject to $\tr \hat{\rho}=1$. Actually, if we choose the generators $\hat{\lambda}_i\ (i=1,\ldots,N^2-1)$ of $SU(N)$ (e.g., the Pauli spin operators in $2$-level system) for observables of interest, the density matrix is determined from their expectation values $\langle \lambda_i \rangle$'s:
\begin{equation}\label{eq:1}
{\bm\lambda} =(\langle \lambda_1 \rangle,\ldots,\langle \lambda_{N^2-1} \rangle) \to \hat{\rho} = \frac{1}{N}\hat{\mathbb{I}}_N + \frac{1}{2}\sum_{i=1}^{N^2-1}\langle \lambda_i \rangle \hat{\lambda}_i.
\end{equation}
(See the next section for details.) Thus, experimentalists have only to measure the values of $\hat{\lambda}_i\ (i=1,\ldots, N^2-1)$ for the identification of a quantum state in the $N$-level system. (If need be, it is possible to determine the corresponding density matrix by using map (\ref{eq:1})). The Bloch vector ${\bm \lambda}$ is defined as ${\bm \lambda} \equiv (\langle \lambda_1 \rangle,\ldots,\langle \lambda_{N^2-1} \rangle) \in \mathbb{R}^{N^2-1}$ and this gives the desirable description of the states for $N$-level systems. However, as concerns the set of the Bloch vectors (the Bloch-vector space) not much has been determined so far. (For experimentalists, the determination of the set is nothing but to prescribe the range of the experimental data observed.) For $N$-level systems, actually the domain constitutes a subset of $\mathbb{R}^{N^2-1}$ but not itself. Notice, however, that not all the vectors ${\bm \lambda} \in \mathbb{R}^{N^2-1}$ give density matrices by map (\ref{eq:1}), because the positivity of $\hat{\rho}$, one of the indispensable properties of the density matrix has yet to be imposed. So far, the complete determination of the domain has been done only in $2$-level systems: It is a ball with radius $1$, known as the Bloch ball. On the other hand, for $N$-level systems ($N\ge 3$), just some properties are known; the Bloch-vector space is a proper subset of a ball in $\mathbb{R}^{N^2-1}$; its $2$-dimensional sections are clarified in $3$-level systems \cite{ref:Mahler} and $4$-level systems ($2$-qubits) \cite{ref:Jakobczyk}; there appears some asymmetric structure \cite{ref:Jakobczyk}; and all of these properties are quite different from those in $2$-level systems. The aim of the present paper is to determine the Bloch-vector space for arbitrary $N$-level systems. We also clarify the origin of its structural difference from the familiar Bloch ball in $2$-level systems.

\section{Review of the Bloch vector for $2$-level systems and preparation for its generalization}

In this section, a brief review of the Bloch vector for $2$-level systems \cite{ref:Bloch,ref:Nielsen} and a preparation for its generalization to $N$-level systems ($N \ge 3$) \cite{ref:HioeEberly,ref:Alicki,ref:Mahler,ref:Jakobczyk,ref:Lendi,ref:Lendi2} are given.
%We will denote ${\mathbb R}^n$ (${\mathbb C}^n$) for the $n$-dimensional real (complex) Euclidian space and $M(n)$ for the $n\times n$ complex matrix space.

\subsection{Density matrix and Generators of $SU(N)$}

We begin with the definition and some properties of the density matrix and generators of $SU(N)$. The density-matrix space $\mathcal{L}_{+,1}(\mathcal{H}_N)$ for $N$-level systems associated with the Hilbert space $\mathcal{H}_N$ ($ \simeq {\mathbb C}^N$) is given by
\begin{eqnarray}\label{eqn:DENSITYMATRIX}
\mathcal{L}_{+,1}(\mathcal{H}_N) = \{ \hat{\rho} \in {\mathcal L}({\mathcal H_N})\ : \mbox{(i)} \ \tr \hat{\rho} =1,\ \mbox{(ii)} \ \hat{\rho} = \hat{\rho}^\dagger , \ \mbox{(iii)} \ \rho_i \ge 0 \  ( i =1,\ldots , N )\},
\end{eqnarray}
where ${\mathcal L}({\mathcal H_N})$ ($ \simeq M(N)$) denotes a set of linear operators on ${\mathcal H}_N$, $\rho_i\ (i=1, \ldots ,N)$ $i$th eigenvalue of $\hat{\rho}$ \cite{note:POSITIVITY}DAs one of the properties of the density matrix,
\begin{equation}\label{eqn:FIDERITY}
\mbox{(iv)} \  \tr \hat{\rho}^2 \le 1
\end{equation}
follows from Eq.~(\ref{eqn:DENSITYMATRIX}) for any $\hat{\rho} \in \mathcal{L}_{+,1}(\mathcal{H}_N)$. Notice that the equality holds if and only if $\hat{\rho}$ is a pure state (i.e., $\exists \ | \psi\rangle  \in {\mathcal H}_N $ s.t. $\hat{\rho} = | \psi\rangle  \langle \psi |$). Condition (iv) (in Eq.~(\ref{eqn:FIDERITY})) plays a role of characterizing the Bloch-vector space to be in a ball in ${\mathbb R}^{N^2-1}$, as will be shown later. In the case of $N=2$, conditions (i), (ii) and (iii) (in Eq.~(\ref{eqn:DENSITYMATRIX})) are equivalent to (i), (ii) and (iv) \cite{note:ProofOf(i)(ii)(iii)<=>(i)(ii)(iv)}:
\begin{eqnarray}\label{eqn:Condition:M(2)}
\forall \hat{\rho} \in {\mathcal L}({\mathcal H_2}),  \
&\mbox{(i)}&\  \tr \hat{\rho} =1, \ \mbox{(ii)} \   \hat{\rho}^\dagger = \hat{\rho}, \ \mbox{(iii)} \ \rho_i \ge 0 \nonumber \\
\Leftrightarrow &\mbox{(i)}&\  \tr \hat{\rho} =1, \ \mbox{(ii)} \   \hat{\rho}^\dagger = \hat{\rho}, \ \mbox{(iv)} \tr \hat{\rho}^2 \le 1.
\end{eqnarray}
This enables us to characterize the density matrix $\hat{\rho}$ for $2$-level systems with condition (i), (ii) and (iv), instead of (iii). Notice that the sufficient condition ($\Leftarrow$) in Eq.~(\ref{eqn:Condition:M(2)}) does not hold for a general case ($N \ge 3$).

The (orthogonal) generators of $SU(N)$ \cite{ref:Mahler} are a set of operators $\hat{\lambda}_i \in {\mathcal L}({\mathcal H}_N) \ (i=1,\ldots, N^2-1)$ which satisfy
\begin{equation}\label{eq:DefOfGeneratorsOfSUn}
\mbox{(a)} \ \hat{\lambda}_i = \hat{\lambda}^\dagger_i, \ \mbox{(b)} \ \tr \hat{\lambda}_i = 0, \ \mbox{(c)} \tr \hat{\lambda}_i\hat{\lambda}_j = 2\delta_{ij}.
\end{equation}
They are characterized with structure constants $f_{ijk}$ (completely antisymmetric tensor) and $g_{ijk}$ (completely symmetric tensor) of Lie algebra $su(N)$:
\begin{subequations}\label{eq:StructureConstants}
\begin{eqnarray}
&[\hat{\lambda}_i,\hat{\lambda}_j]&  = 2i f_{ijk}\hat{\lambda}_k \\
{}&[\hat{\lambda}_i,\hat{\lambda}_j]_+& = \frac{4}{N}\delta_{ij}\hat{\mathbb{I}}_N + 2 g_{ijk}\hat{\lambda}_k,
\end{eqnarray}
\end{subequations}
where $[\cdot,\cdot]$ and $[\cdot,\cdot]_+$ are respectively a commutation and an anticommutation relation. (Repeated indices are summed from $1$ to $N^2-1$.)

A systematic construction of generators of $SU(N)$ which generalize the Pauli spin operators is known \cite{ref:HioeEberly,ref:Mahler,ref:Alicki}; the (orthogonal) generators are given by
\begin{equation}\label{eqn:SpecificGeneratorSU(n)}
\{ \hat{\lambda}_i \}_{i=1}^{N^2-1} = \{ \hat{u}_{jk}, \hat{v}_{jk}, \hat{w}_{l}  \},
\end{equation}
where
\begin{subequations}
\begin{eqnarray}
\hat{u}_{jk} &=& |j \rangle \langle k | + |k \rangle \langle j |, \ \hat{v}_{jk} = -i(|j \rangle \langle k | - |k \rangle \langle j |), \\
\hat{w}_{l} &=&  \sqrt{\frac{2}{l(l+1)}}\sum_{j=1}^{l} (|j \rangle \langle j | - l|l+1 \rangle \langle l+1 |),    \\
&&1 \le j \le k \le N, \ 1 \le l \le N-1,
\end{eqnarray}
\end{subequations}
with $\{ |m \rangle \}^{N}_{m=1}$ being some complete orthonormal basis of ${\mathcal H}_N$. This gives Pauli spin operators ($\hat{\lambda}_1  = \hat{u}_{12}\equiv \hat{\sigma}_1, \ \hat{\lambda}_2 = \hat{v}_{12}\equiv \hat{\sigma}_2, \ \hat{\lambda}_3 =\hat{w}_1\equiv \hat{\sigma}_3$) for $N=2$ with structure constants
\begin{equation}\label{eq:StructureConstantsFor2}
f_{ijk} = \epsilon_{ijk}\  \mbox{(Levi-Civita symbol)}, \ g_{ijk}=0,
\end{equation}
and Gell-Mann operators ($\hat{\lambda}_1 = \hat{u}_{12}, \ \hat{\lambda}_2 = \hat{v}_{12}, \ \hat{\lambda}_3 =\hat{w}_1, \ \hat{\lambda}_4 =\hat{u}_{13}, \ \hat{\lambda}_5 =\hat{v}_{13}, \ \hat{\lambda}_6 =\hat{u}_{23}, \ \hat{\lambda}_7 =\hat{v}_{23}, \ \hat{\lambda}_8 =\hat{w}_{2}$) for $N=3$ with non-vanishing structure constants:
\begin{eqnarray}\label{eqn:DefiningStructureConstants}
f_{123} &=& 1,\nonumber \\
 f_{458} &=& f_{678} = \sqrt{3}/2,\nonumber \\
 f_{147} &=& f_{246} = f_{257} =f_{345} = -f_{156} =-f_{367} = 1/2, \nonumber \\
g_{118} &=& g_{228} = g_{338} = -g_{888} = \sqrt{3}/3,\nonumber \\
g_{448} &=& g_{558} = g_{668} = g_{778} = -\sqrt{3}/6, \nonumber \\
g_{146} &=& g_{157} = g_{256} = g_{344} = g_{355} = -g_{247} =-g_{366}= -g_{377} =  1/2.
\end{eqnarray}
 Any generators $\{\hat{\lambda^{\prime}}_i\}_{i=1}^{N^2-1}$ which satisfy Eqs.~(\ref{eq:DefOfGeneratorsOfSUn}) with structure constants $f^{\prime}_{ijk},g^{\prime}_{ijk}$ are connected with this specific generators (\ref{eqn:SpecificGeneratorSU(n)}) by some orthogonal matrix $V \in O(N^2-1)$ with $\hat{\lambda^{\prime}}_i = V_{ij}\hat{\lambda}_j$ and
\begin{equation}\label{eq:TransformationOfFandG}
f^{\prime}_{ijk} = V_{il}V_{jm}V_{kn}f_{lmn}, \ g^{\prime}_{ijk} = V_{il}V_{jm}V_{kn}g_{lmn}.
\end{equation}
Since the Levi-Civita tensor has a rotational invariance, i.e., $V_{il}V_{jm}V_{kn}\epsilon_{ijk} = \det V \epsilon_{ijk} = \pm \epsilon_{ijk}$, the structure constants of $SU(2)$ are limited to $f_{ijk} = \pm \epsilon_{ijk}$ and $ \ g_{ijk} = 0$, while in the case of $N \ge 3$ there is no rotational invariance of structure constants.

Finally we notice that the generators $\hat{\lambda}_i$'s of $SU(N)$ with an identity operator $\hat{{\mathbb I}}_N \in {\mathcal L}({\mathcal H}_N)$ form a complete orthogonal basis of ${\mathcal L}({\mathcal H}_N)$, in the sense of the Hilbert-Schmidt product.

\subsection{The Bloch vector for $2$-level systems}

We review the familiar Bloch vector for $2$-level systems ($N=2$). From the properties (\ref{eq:DefOfGeneratorsOfSUn}) for $N=2$, it is easy to show that the following statements $(\mbox{I}_2)$ and $(\mbox{II}_2)$ hold:

$(\mbox{I}_2)$ Any operator $\hat{\rho} \in {\mathcal L}({\mathcal H}_2)$ with conditions (i) and (ii) can be uniquely characterized by a $3$-dimensional real vector ${\bm \lambda} = (\lambda_1,\lambda_2,\lambda_3) \in {\mathbb R}^3$ as
\begin{equation}
\hat{\rho}=\frac{1}{2}\hat{{\mathbb I}}_2 + \frac{1}{2} \lambda_i \hat{\sigma}_i.
\end{equation}

$(\mbox{II}_2)$ By imposing condition (iv) on the above operator $\hat{\rho}$, the length of ${\bm \lambda}$ is restricted to be less than or equal to $1$:
\begin{equation}\label{eq:SuperBall2}
| {\bm \lambda}| \equiv \sqrt{\lambda_i\lambda_i} \le 1.
\end{equation}

\begin{figure}[]
\includegraphics[height=0.4\textwidth]{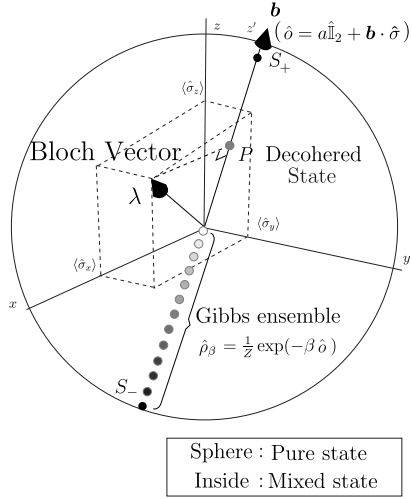}
\caption{Bloch ball in $2$-level systems}
\label{fig:BlochBall}
\end{figure}

From these statements and relation (\ref{eqn:Condition:M(2)}), any density matrix in $2$-level systems turns out to be characterized uniquely by a $3$-dimensional real vector where the length satisfies Eq.~(\ref{eq:SuperBall2}). Therefore, if we define the Bloch-vector space $B({\mathbb R}^3)$ as a ball with radius $1$:
\begin{equation}\label{eq:BlochBall}
B({\mathbb R}^3) = \{ \ {\bm \lambda}= (\lambda_1,\lambda_2,\lambda_3) \in {\mathbb R}^3 \ : \ |{\bm \lambda}| \le 1 \ \},
\end{equation}
its element gives an equivalent description of the density matrix with the following bijection (one-to-one and onto) map from $B({\mathbb R}^3)$ to $\mathcal{L}_{+,1}(\mathcal{H}_2) $:
\begin{equation}\label{eqn:MapBtoT}
{\bm \lambda}   \to \hat{\rho} = \frac{1}{2}\hat{{\mathbb I}}_2 + \frac{1}{2} \lambda_i \hat{\sigma}_i.
\end{equation}
$B({\mathbb R}^3)$ is called the Bloch ball, its surface the Bloch sphere and its element the Bloch vector. Since the equality in Eq.~(\ref{eq:SuperBall2}) (i.e., $|{\bm \lambda}| = 1$) originates from the one in condition (iv), the surface of the ball (the Bloch sphere) corresponds to the set of pure states and its inside to mixed states. The inverse map of (\ref{eqn:MapBtoT}) is
\begin{equation}
\hat{\rho} \to \lambda_i = \tr \hat{\rho} \hat{\sigma}_i \quad (i=1,2,3).
\end{equation}
Since we can consider $\hat{\sigma}_i$'s to be some observables, the physical meaning of a component of the Bloch vector is an expectation value of $\hat{\sigma}_i$: ${\bm \lambda} = (\langle \sigma_1 \rangle,\langle \sigma_{2} \rangle,\langle \sigma_{3} \rangle)$. An important point is that a state can be characterized by expectation values of $\hat{\sigma}_i$'s which are directly observed in experiments. Furthermore, the Bloch vectors allow us to grasp characteristics of states from a completely geometrical stand point: The components of the Bloch vector themselves give expectation values of $\hat{\sigma}_i$'s. In addition, for any observable $\hat{o} = a \hat{{\mathbb I}}_2 + b_i \hat{\sigma}_i \ (a,\  b_i \in {\mathbb R}$), its expectation value is $a$ plus {\it the new $z^{\prime}$-component} of ${\bm \lambda}$ multiplied by $|{\bm b}|$ if we consider the direction of ${\bm b}$ as the new $z^{\prime}$-axis (See Fig.~\ref{fig:BlochBall}); the probabilities to observe eigenvalues $a+|{\bm b}|$ and $a-|{\bm b}|$ of $\hat{o}$ correspond to the normalized rates of $P-S_{-}$ and $P-S_{+}$ respectively, where $P$ is the projective point from ${\bm \lambda}$ to a direction of ${\bm b}$ and $S_{+}$ ($S_{-}$) the points on the surface of a positive (negative) direction of ${\bm b}$. The specific states such as the Gibbs state are also characterized in $B({\mathbb R}^3)$; the Gibbs state $\hat{\rho}_{\beta} = \frac{1}{Z}\exp(-\beta \hat{H})$ with Hamiltonian $\hat{H} = \hat{o}$ and temperature $1/\beta$ corresponds to a line segment between origin $O$ (infinite temperature : $1/\beta = \infty$) and $S_-$ (zero temperature : $1/\beta = 0$). The decohered state after observation of $\hat{o}$ corresponds to point $P$. The time evolution of a state can also be captured as an orbit in the Bloch ball and it gives a clear visualization for the Larmor precession process, thermalization process or decoherence process and other time evolutions.

\subsection{Preparation for the generalization of the Bloch vector}\label{sec:A preparation for generalizing of a Bloch vector}

The generalization of the Bloch vector to $N$-level systems ($N\ge 3$) can be done similarly to the case of $2$-level systems to some extent. If we choose the generators of $SU(N)$ as the observables of interest, then the same statements $(\mbox{I}_N)$ and $(\mbox{II}_N)$ for $N$-level systems hold from properties (\ref{eq:DefOfGeneratorsOfSUn}):

$(\mbox{I}_N)$ Any operator $\hat{\rho} \in {\mathcal L}({\mathcal H}_N)$ with conditions (i) and (ii) can be uniquely characterized by a ($N^2-1$)-dimensional real vector ${\bm \lambda} = (\lambda_1,\ldots,\lambda_{N^2-1}) \in {\mathbb R}^{N^2-1}$ as
\begin{equation}\label{eq:FormOfBlochVector}
\hat{\rho}=\frac{1}{N}\hat{{\mathbb I}}_N + \frac{1}{2} \lambda_i \hat{\lambda}_i.
\end{equation}

$(\mbox{II}_N)$ By imposing condition (iv) on the above operator $\hat{\rho}$, the length of ${\bm \lambda}$ is restricted to be less than or equal to $\sqrt{2(N-1)/N}$:
\begin{equation}\label{eq:SuperBalln}
| {\bm \lambda}| \equiv \sqrt{\lambda_i\lambda_i} \le \sqrt{\frac{2(N-1)}{N}}.
\end{equation}
However this does not complete the generalization of the Bloch vector, since properties (i), (ii) and (iv) are only necessary conditions for (i), (ii) and (iii) (i.e., to be the density matrix), but not sufficient for $N\ge 3$. This implies that the Bloch-vector space $B({\mathbb R}^{N^2-1})$ ($N\ge 3$) is a proper subset of a ball (\ref{eq:SuperBalln}). Actually, it has been shown \cite{ref:Jakobczyk} that an angle between any two Bloch vectors ${\bm \lambda}_\alpha,{\bm \lambda}_\beta$ satisfies
\begin{equation}
1 \ge \cos \angle(\alpha,\beta) \ge -\frac{1}{N-1},
\end{equation}
if the corresponding density matrices $\hat{\rho}_\alpha$ and $\hat{\rho}_\beta$ are pure, i.e., $\hat{\rho}_\alpha^2 = \hat{\rho}_\alpha, \hat{\rho}_\beta^2 = \hat{\rho}_\beta$, from the following property
\begin{equation}
1 \ge \tr \hat{\rho}_\alpha \hat{\rho}_\beta \ge 0.
\end{equation}
This shows $B({\mathbb R}^{N^2-1})$ occupies a relatively small part of the ball and implies an asymmetric structure for the case of $N\ge 3$. Furthermore, $2$-dimensional sections of $B({\mathbb R}^{N^2-1})$ for $N=3$ \cite{ref:Mahler} and $N=4$ ($2$-qubits) \cite{ref:Jakobczyk} are examined in detail, which also support these features. However, the complete determination of the Bloch-vector space for an arbitrary dimension has not been done. In the next section, we will completely specify it for any $N$-level systems.

\section{The Bloch vector for $N$-level systems}\label{sec:Bloch vector in finite level systems}

It is clear that a characterization of the Bloch vector depends on how to integrate condition (iii) (instead of (iv)) into the form of Eq.~(\ref{eq:FormOfBlochVector}). For this purpose, the following lemma will be useful.
\begin{lemma}\label{lem:POSITIVITY}
Consider an algebraic equation of degree $N \ge 1$:
\begin{equation}\label{eqn:AlgebraicEquation}
\sum_{j=0}^{N}(-1)^j a_j x^{N-j}= \prod_{i=1}^{N}(x-x_i) = 0\ (a_0 = 1),
\end{equation}
which has only real roots $x_i \in {\mathbb R}\ (i=1,\ldots,N)$. The necessary and sufficient condition that all the roots $x_i$'s to be positive semi-definite is that all the coefficients $a_i$'s are positive semi-definite:
\begin{equation}
x_i \ge 0 \ (i=1,\ldots,N) \Leftrightarrow a_i \ge 0 \ (i=1,\ldots,N).
\end{equation}
\end{lemma}
This lemma can be considered as a corollary of the famous Descartes' theorem (rule's of signs) \cite{ref:DescartesAndNewton}. (For the reader's convenience, we give a direct proof of Lemma \ref{lem:POSITIVITY} in Appendix \ref{ProofofLemma1}). From this lemma we obtain
\begin{theorem}\label{thm:Sigma:n}

Let $a_i({\bm \lambda})$'s be coefficients of the characteristic polynomial $\det (x \hat{{\mathbb I}}_N  - \hat{\rho})$ where $\hat{\rho}$ is an operator of the form (\ref{eq:FormOfBlochVector}) and define
\begin{equation}\label{eq:B:n:CONDITION}
B({\mathbb R}^{N^2-1}) = \{ {\bm \lambda} \in {\mathbb R}^{N^2-1}  :  a_i({\bm \lambda}) \ge 0 \ (i=1,\ldots,N)\}.
\end{equation}
Then a map:
\begin{equation}\label{eqn:MapBtoTgeneral:n}
{\bm \lambda} \in B({\mathbb R}^{N^2-1}) \to \hat{\rho} = \frac{1}{n}\hat{{\mathbb I}}_N + \frac{1}{2} \lambda_i \hat{\lambda}_i \in {\mathcal L}({\mathcal H}_N)
\end{equation}
is a bijection from $B({\mathbb R}^{N^2-1})$ to the density-matrix space ${\mathcal L}_{+,1}({\mathcal H}_N)$.
\end{theorem}
{\bf Proof of Theorem \ref{thm:Sigma:n}}

The operator $\hat{\rho}$ in Eq.~(\ref{eqn:MapBtoTgeneral:n}) clearly satisfies conditions (i) and (ii) ({\it Hermiticity}) from $(\mbox{I}_N)$. By taking account of a fact that an Hermitian operator has a real spectrum, the eigenvalues $\rho_i$'s of $\hat{\rho}$ turn out to be all positive semi-definite, i.e., condition (iii) holds from the sufficient condition of Lemma \ref{lem:POSITIVITY} and the conditions $a_i({\bm \lambda}) \ge 0 \ (i=1,\ldots,N)$ in Eq.~(\ref{eq:B:n:CONDITION}). Therefore, the operator is a density matrix and Eq.~(\ref{eqn:MapBtoTgeneral:n}) is a map from $B({\mathbb R}^{N^2-1})$ to ${\mathcal L}_{+,1}({\mathcal H}_N)$. The injectivity (one-to-one property) comes from linear independence of $\hat{\mathbb I}_N$ and $\hat{\lambda}_i$'s; the surjectivity (onto property) comes from the necessary condition of Lemma \ref{lem:POSITIVITY}.

 \hfill{Q.E.D}

Theorem \ref{thm:Sigma:n} states that the Bloch-vector space for $N$-level systems is nothing but $B({\mathbb R}^{N^2-1})$ in Eq.~(\ref{eq:B:n:CONDITION}) with the bijection map (\ref{eqn:MapBtoTgeneral:n}), which gives the corresponding density matrix. Components of the Bloch vector can be considered as expectation values of $\hat{\lambda}_i$'s
\begin{equation}
\lambda_i = \tr \hat{\rho} \hat{\lambda}_i \quad (i=1,\ldots,N^2-1),
\end{equation}
which give the inverse map of (\ref{eqn:MapBtoTgeneral:n}). This means that the state of $N$-level systems can be completely characterized with ($N^2-1$)-expectation values of observables $\hat{\lambda}_i$'s.

Notice that the coefficients $a_i({\bm \lambda})\ (i=1,\ldots,N)$ can be written down explicitly if we take a matrix representation of the operator $\hat{{\mathbb I}}_N/N + \lambda_i \hat{\lambda}_i/2$. Therefore, the expression of $B({\mathbb R}^{N^2-1})$ in Eq.~(\ref{eq:B:n:CONDITION}) is a practical description of the Bloch-vector space. However, it will be further convenient and instructive to express the conditions $a_i({\bm \lambda})\ge 0 \ (i=1,\ldots,N)$ without resort to particular matrix representation. Such an expression clarifies the relation to the previous discussion in Sec.~\ref{sec:A preparation for generalizing of a Bloch vector} and the difference between $N=2$ and $N\ge 3$ cases. To obtain it, we use the famous Newton's formulas \cite{ref:DescartesAndNewton} which connect coefficients $a_i\ (i=1,\ldots,N)$ ($a_0=1$) and the sums of the powers of roots $x_i\ (i=1,\ldots,N)$ of the algebraic equation of $N$ degrees (\ref{eqn:AlgebraicEquation}): Newton's formulas reads
\begin{subequations}\label{eq:ExplicitNewtonFormula}
\begin{equation}
ka_k = \sum_{q=1}^{k}(-1)^{q-1}C_{N,q}a_{k-q} \quad (1\le k \le N),\end{equation}
where $C_{N,q} \equiv \sum_{i=1}^{N}x^{q}_i$. Explicitly,
\begin{eqnarray}
a_1 &=& C_{1},\nonumber \\
a_2 &=& (C_{1}^2 - C_{2})/2, \nonumber \\
a_3 &=& (C_{1}^3 -3C_{1}C_{2}+2C_{3})/3!, \nonumber \\
a_4 &=& (C_{1}^4-6C_{1}^2C_{2}+8C_{1}C_{3}+3C_{2}^2-6C_{4})/4!, \nonumber \\
  a_5 &=& \cdots,
\end{eqnarray}
\end{subequations}
where $C_{N,q}$ is simply denoted as $C_{q}$ here. Applying this to the characteristic polynomial of $\hat{\rho}$ of the form (\ref{eq:FormOfBlochVector}), one obtains
\begin{subequations}
\begin{eqnarray}
1!a_1 &=& 1, \label{eqn:a1=1}\\
2! a_2 &=& 1-\tr \hat{\rho}^2 \label{eqn:a2=1-TrRho^2},
\end{eqnarray}
and
\begin{eqnarray}
3! a_3 &=& 1 -3\tr \hat{\rho}^2+2\tr \hat{\rho}^3, \nonumber \\
4! a_4 &=& 1-6\tr \hat{\rho}^2+8\tr \hat{\rho}^3+3(\tr \hat{\rho}^2)^2-6\tr \hat{\rho}^4, \nonumber\\
5! a_5 &=& \cdots,
\end{eqnarray}
\end{subequations}
where use has been made of $C_{N,q} = \tr \hat{\rho}^q$ and $C_{N,1} = \tr \hat{\rho} = 1$. Equation (\ref{eqn:a1=1}) means that the condition $a_1({\bm \lambda}) \ge 0$ trivially holds, while Eq.~(\ref{eqn:a2=1-TrRho^2}) tells us that the condition $a_2({\bm \lambda}) \ge 0$ is equivalent to condition (iv) in Eq.~(\ref{eqn:FIDERITY}). The latter has made the Bloch-vector space be in a ball (\ref{eq:SuperBalln}). (See statement $(\mbox{II}_N)$ in Sec.~\ref{sec:A preparation for generalizing of a Bloch vector}.) Consequently we understand that while in $2$-level systems the Bloch-vector space is exactly a ball itself because the coefficients exist up to $a_2$, in $N$-level systems ($N\ge 3$) there are additional conditions $a_3 \ge 0, a_4 \ge 0, \cdots, a_N \ge 0$, which restrict the Bloch-vector space to be a proper subset of a ball.

One can further obtain the concrete expressions of coefficients $a_i({\bm \lambda})$'s in Eq.~(\ref{eq:B:n:CONDITION}) in terms of the structure constants:
\begin{eqnarray}\label{eq:Concrete:ai}
&1! a_1& = 1, \nonumber \\
&2! a_2& =(\frac{N-1}{N}-\frac{1}{2}|{\bm \lambda}|^2),\nonumber \\
&3! a_3& =\Big[\frac{(N-1)(N-2)}{N^2} - \frac{3(N-2)}{2N}|\lambda|^2 + \frac{1}{2}g_{ijk}\lambda_i\lambda_j\lambda_k\Big], \nonumber\\
&4!a_4 & = \Big[\frac{(N-1)(N-2)(N-3)}{N^3}-\frac{3(N-2)(N-3)}{N^2}|\lambda|^2 +\frac{3(N-2)}{4N}|\lambda|^4 \nonumber \\
&&\qquad {} +\frac{2(N-2)}{N}g_{ijk}\lambda_i\lambda_j\lambda_k - \frac{3}{4}g_{ijk}g_{klm}\lambda_i\lambda_j\lambda_l\lambda_m\Big] \nonumber \\
&5!a_5 & = \cdots,
\end{eqnarray}
where the completely antisymmetric property of $f_{ijk}$ has been taken into account to evaluate the terms in which $f_{ijk}$ appears. See Appendix \ref{CalculationsForCondition} for the detailed calculations. (Notice that $a_i$'s in Eqs.~(\ref{eq:Concrete:ai}) have meaning only for $i\le N$.) It deserves to say that since the structure constants $g_{ijk}$ of $SU(N)\ (N\ge 3)$ have no rotational invariance, neither do these conditions. Thus the Bloch-vector space has an asymmetric structure in ${\mathbb R}^{N^2-1}$ for $N\ge 3$.
%Instead, they have an invariance in changing any two subscripts because of the complete symmetric property of $g_{ijk}$.

In the following, we illustrate the Bloch-vector space for $3$-level systems, the simplest but non-trivial case. For $3$-level systems, Eqs.~(\ref{eq:Concrete:ai}) read: $a_1 = 1, \ 2! a_2 =(\frac{2}{3}-\frac{1}{2}|{\bm \lambda}|^2)$, and
\begin{equation}\label{eq:a3}
3! a_3 =\Big[\frac{2}{9} - \frac{1}{2}|{\bm \lambda}|^2 + \frac{1}{2}g_{ijk}\lambda_i\lambda_j\lambda_k\Big],
\end{equation}
and the Bloch-vector space (\ref{eq:B:n:CONDITION}) is a ball in ${\mathbb R}^8$ with radius $2/\sqrt{3}$, subject to an additional condition
\begin{equation}\label{eq:AditionalConsition3}
36 - 9|{\bm \lambda}|^2 + 9g_{ijk}\lambda_i\lambda_j\lambda_k \ge 0.
\end{equation}
By using the explicit values of the structure constants for the specific generators (\ref{eqn:SpecificGeneratorSU(n)}), the condition (\ref{eq:AditionalConsition3}) ($a_3 \ge 0$) reads
\begin{eqnarray}\label{eq:AditionalConsition3:Specific}
-8 + 18|{\bm \lambda}|^2
 - 27\lambda_3(\lambda^2_4+\lambda^2_5 - \lambda^2_6-\lambda^2_7)+6\sqrt{3}\lambda^3_8\nonumber \\
{}-9\sqrt{3}\left\{2(\lambda^2_1+\lambda^2_2+\lambda^2_3) - (\lambda^2_4+\lambda^2_5+\lambda^2_6+\lambda^2_7)\right\}\nonumber \\
{}- 54(\lambda_1\lambda_4\lambda_6+\lambda_1\lambda_5\lambda_7+ \lambda_2\lambda_5\lambda_6-\lambda_2\lambda_4\lambda_7) \ge 0.
\end{eqnarray}

In order to visualize the Bloch-vector space, we use $2$-dimensional sections $\Sigma_2(i,j)$ \cite{ref:Jakobczyk,ref:Mahler} which are defined as $\Sigma_2(i,j) = \{ {\bm \lambda} \in B({\mathbb R}^8) : {\bm \lambda} = (0,\ldots,0,\lambda_i,0,\ldots,0,\lambda_j,0,\ldots,0)\}$. In the $2$-dimensional sections, Eq.~(\ref{eq:AditionalConsition3:Specific}) are classified into $4$ types of conditions:

%\begin{widetext}
\begin{figure}
\includegraphics[height=0.275\textwidth]{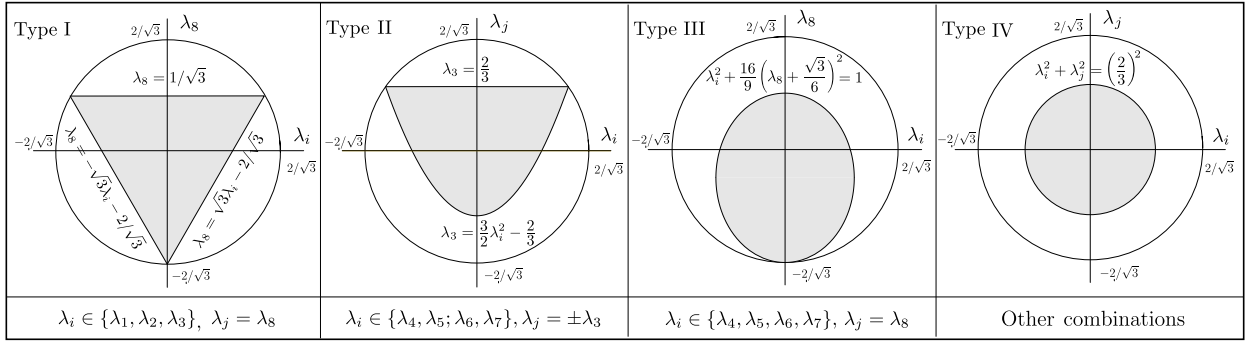}
\caption{
$2$-dimensional sections of the Bloch-vector space for $3$-level systems: Large circles are sections of the ball $|{\bm \lambda}| \le 2/\sqrt{3}$. Grey parts are the domain of the Bloch vector.
}
\label{fig:Section}
\end{figure}
%\end{widetext}

Type I: Where $\lambda_i \in \{ \lambda_1,\lambda_2,\lambda_3 \}$ and $\lambda_j = \lambda_8$, then
\begin{equation}
\lambda_8  \le 1/\sqrt{3},\  \lambda_8 \ge \pm \sqrt{3}\lambda_i - 2/\sqrt{3}.
\end{equation}

Type II: Where $ \lambda_i \in \{ \lambda_4,\lambda_5 \}$ and $ \lambda_j= \lambda_3$ or $\lambda_i \in \{ \lambda_6,\lambda_7 \}$ and $\lambda_j= - \lambda_3$, then
\begin{equation}
\lambda_j  \le \frac{2}{3},\  \lambda_j \ge \frac{3}{2}\lambda^2_i-\frac{2}{3}.
\end{equation}

Type III: Where $ \lambda_i \in \{ \lambda_4,\lambda_5,\lambda_6,\lambda_7 \}$ and $ \lambda_j= \lambda_8$, then
\begin{equation}
\lambda^2_i + \frac{16}{9}\left(\lambda_8 + \frac{\sqrt{3}}{6}\right)^2 \le 1.
\end{equation}

Type IV: Otherwise,
\begin{equation}
\lambda^2_i + \lambda^2_j \le \left(\frac{2}{3}\right)^2.
\end{equation}

Combined with the ball condition $|{\bm \lambda}| \le 2/\sqrt{3}$, we illustrate the $4$ types of sections of the Bloch-vector space in Fig.~\ref{fig:Section}. It is clear that the Bloch-vector space is a proper subset of the ball. In addition, one sees quite asymmetric structures for Types I, II and III, which stem from the absence of the rotational invariance of the generators $g_{ijk}$ in the condition (\ref{eq:AditionalConsition3}).

\section{Conclusion and discussion}

We have characterized the Bloch-vector space for arbitrary $N$-level systems as that prescribed by Eq.~(\ref{eq:B:n:CONDITION}); and their explicit expressions are given in Eqs.~(\ref{eq:Concrete:ai}). The essential difference between $2$-level and $N$-level systems ($N\ge 3$) is whether they have conditions up to $a_2$ (the ball condition) or more (i.e., $a_3 \ge 0, a_4 \ge 0, \cdots a_N \ge 0$). Asymetric structures appear in $N$-level systems ($N\ge 3$) because the structure constants $g_{ijk}$ have no rotational invariance.

The classification of the states such as pure or mixed states, separable or entangled states \cite{ref:Werner,ref:Nielsen} can be also arranged by means of the Bloch vector: The pure states correspond to the surface of the ball (\ref{eq:SuperBalln}) and mixed states inside. As concerns the separability and entanglement \cite{ref:Jakobczyk,ref:Entanglement,ref:Zyczkowski,ref:Braunstein}, we can use the famous Peres' criterion \cite{ref:Peres} (positive partial transpose) in $2\times 2$ or $2 \times 3$ composite systems \cite{ref:Horodecki}. Using Lemma \ref{lem:POSITIVITY} to check the positivity in the partially transposed transformation, we can determine the sets of separable and entangled states in the Bloch-vector space \cite{ref:Hayashi}.

Although the Bloch-vector space is completely specified, it does not mean that we have established a natural parameterization \cite{ref:Tilma} like in $2$-level systems. In $2$-level systems, it is nothing but the Bloch ball, which can be naturally parameterized by the polar coordinates. Considering the complex structures (brought about by the remaining constraints $a_3 \ge 0, a_4 \ge 0, \cdots$) in $N$-level systems ($N\ge 3$), the notion of the Bloch vector in higher dimensional systems might not be as useful as in $2$-level systems. However it is still considered to be important to express quantum states in terms of the expectation values of observables, since then experimentalists can directly determine the states with the use of their experimental results. In the circumstances, the classification of the state in the Bloch-vector space is further meaningful so that they can find whether the state is pure or mixed, separable or entangled, etc., with their data.

%---------- ACKNOWLEDGEMENT -------------------------------------
\acknowledgments

The author would like to thank Prof. I. Ohba, Prof. S. Tasaki, Dr. K. Imafuku, Dr. T. Hirano and Dr. T. Tilma for their valuable comments and advice. He is grateful to Prof. H. Nakazato and Dr. Y. Ota for reading the manuscript prior publication and fruitful discussion. He also thanks Dr. M. Miyamoto and Dr. S. Mine for valuable comments and advice on mathematical aspects of the analysis.

After the completion of this work, we have noticed a preprint of a related work by M. S. Byrd and N. Khaneja \cite{ref:Byrd}, in which the same results of Eqs.~(\ref{eq:Concrete:ai}) were derived; the application to the composite systems is given in detail.

\appendix

\section{Proof of Lemma \ref{lem:POSITIVITY}}\label{ProofofLemma1}

We present a direct proof of Lemma \ref{lem:POSITIVITY}.

\noindent{\bf Proof of Lemma \ref{lem:POSITIVITY}}

\noindent [Necessary condition]F From Vieta's formula which connects roots and coefficients:
\begin{equation}
a_i = \sum_{1\le j_1<j_2<\cdots<j_i}^{N} x_{j_1}x_{j_2}\cdots x_{j_i}\ (i=1,\ldots,N)
\end{equation}
it is clear that $x_i \ge 0 \ (i=1,\ldots,N) \Rightarrow a_i \ge 0 \ (i=1,\ldots,N)$.

\noindent [Sufficient condition]FLet $a_i \ge 0 \ (i=1,\ldots,N)$ and assume at least one of the roots is negative definite. (Without loss of generality, we can put $x_N < 0$). Let us define $\{\tilde{a}_i\}_{i=1}^{N-1}$ as:
\begin{equation}
\sum_{j=0}^{N}(-1)^j a_j x^{N-j} = (x-x_N) \sum_{j=0}^{N-1}(-1)^j \tilde{a}_j x^{(N-1)-j} \ (\tilde{a}_0 = 1),
\end{equation}
then clearly the following relations
\begin{equation}\label{eqn:ZENKA}
a_i = \tilde{a}_i + \tilde{a}_{i-1}x_N\  (i=1,\ldots N),
\end{equation}
in which $\tilde{a}_N = 0$ hold. In the case of $i=N$ in (\ref{eqn:ZENKA}), it follows that $\tilde{a}_{N-1} \le 0$ because $a_{N} \ge 0, \ \tilde{a}_{N} = 0$ and $ x_{N} < 0$. In the case of $i=N-1$ in (\ref{eqn:ZENKA}), it follows that $\tilde{a}_{N-2} \le 0$ because $a_{N-1} \ge 0, \ \tilde{a}_{N-1} \le 0$ and $x_{N} < 0$. Continuing this deduction successively for $i=N-1, N-2, \cdots$ in (\ref{eqn:ZENKA}), it follows that $\tilde{a}_{1} \le 0$ for $i=2$ and finally we obtain $a_1 = \tilde{a}_1 + \tilde{a}_0 x_N < 0$. However this contradicts one of the assumptions $a_1 \ge 0$.
\hfill{QED}

\section{Calculations of $a_i$'s in Eq.~(\ref{eq:Concrete:ai})}\label{CalculationsForCondition}

From Eqs.~(\ref{eq:DefOfGeneratorsOfSUn}) and (\ref{eq:StructureConstants}), one obtains
\begin{eqnarray}
&&\tr\hat{\lambda}_i = 0,\ \tr\hat{\lambda}_i\hat{\lambda}_j = 2\delta_{ij}, \ \tr \hat{\lambda}_i\hat{\lambda}_j\hat{\lambda}_k = 2 z_{ijk},\nonumber \\
&&\tr \hat{\lambda}_i\hat{\lambda}_j\hat{\lambda}_k\hat{\lambda}_l = \frac{4}{N}\delta_{ij}\delta_{kl} + 2 z_{ijm}z_{mkl}, \nonumber \\
&& \tr \hat{\lambda}_i\hat{\lambda}_j\hat{\lambda}_k\hat{\lambda}_l\hat{\lambda}_m = \frac{4}{N}\delta_{ij}z_{klm} + \frac{4}{N}\delta_{lm}z_{ijk} + 2 z_{ijn}z_{nko}z_{olm}\nonumber \\
&& \tr \hat{\lambda}_i\hat{\lambda}_j\hat{\lambda}_k\hat{\lambda}_l\hat{\lambda}_m\hat{\lambda}_n = \cdots,
\end{eqnarray}
in which $z_{ijk} \equiv g_{ijk}+if_{ijk}$. Then it is easy to calculate $C_{N,q}= \tr \hat{\rho}^q$ where $\hat{\rho}$ is of the form (\ref{eq:FormOfBlochVector}):
\begin{eqnarray}
C_{N,1} &=&  1, \nonumber \\
C_{N,2} &=&  \frac{1}{(2N)^2} (4N+ 2N^2 |{\bm \lambda}|^2), \nonumber \\
C_{N,3} &=& \frac{1}{(2N)^3}(8N + 12N^2|{\bm \lambda}|^2 + 2N^3 \lambda_i\lambda_j\lambda_k g_{ijk}), \nonumber \\
C_{N,4} &=& \frac{1}{(2N)^4}\Big(16N + 48N^2|{\bm \lambda}|^2 + 16N^3 \lambda_i\lambda_j\lambda_k g_{ijk} +4N^3 |{\bm \lambda}|^4 + 2N^4 \lambda_i\lambda_j\lambda_k\lambda_l g_{ijm}g_{mkl} \Big), \nonumber \\
C_{N,5} &=&  \cdots.
\end{eqnarray}
Substituting these expressions for $C_{N,q}$'s in Newton's formulas (\ref{eq:ExplicitNewtonFormula}), we obtain the explicit expressions in Eq.~(\ref{eq:Concrete:ai}).

%---------- REFERENCES ------------------------------------------


\begin{thebibliography}{10}
\bibitem{ref:Weigert2}
For the determination of states, ($N^2-1$)-observables are usually not necessary, because one can extract more information from one observable (e.g., by its probability distribution) \cite{ref:Weigert}.
\bibitem{note:POSITIVITY}
Notice that the conditions $\mbox{(ii)}$ and $\mbox{(iii)}$ are nothing but the positivity of the density matrix $\hat{\rho} \ge 0$. It is rather convenient to divide it into $\mbox{(ii)}$ and $\mbox{(iii)}$ when we discuss the Bloch vector.
\bibitem{note:ProofOf(i)(ii)(iii)<=>(i)(ii)(iv)}
As a nontrivial case, suppose that (i) and (ii) hold but (iii) does not. Let $\rho_1= -\epsilon \ (\epsilon > 0)$ be a negative-definite eigenvalue of $\hat{\rho}$, and $\rho_2 = 1 + \epsilon$, then $\tr \hat{\rho}^2 = 1 + 2\epsilon^2 + 2\epsilon > 1$ holds. This means (iv) does not hold.
\bibitem{ref:Entanglement}
The question of how many separable states there are is considered in Refs.~\cite{ref:Zyczkowski} and \cite{ref:Braunstein} using a suitable measure for the quantum states. An interesting dualism between purity and separability is also discussed.
\end{thebibliography}
\end{document}